\documentclass[runningheads]{llncs}
\usepackage[utf8]{inputenc}
\usepackage{graphicx}
\usepackage{caption}
\usepackage{float}
\usepackage[backgroundcolor=blue!10,bordercolor=gray]{todonotes}

\graphicspath{{fig/}}

\begin{document}
\title{A Data-Mining Based Study of Security Vulnerability Types and their Mitigation in Different Languages\thanks{The presented work was carried out within the SETIT Project (2018-1.2.1-NKP-2018-00004). Project no. 2018-1.2.1-NKP-2018-00004 has been implemented with the support provided from the National Research, Development and Innovation Fund of Hungary, financed under the 2018-1.2.1-NKP funding scheme and partially supported by grant TUDFO/47138-1/2019-ITM of the Ministry for Innovation and Technology, Hungary.
Furthermore, Péter Hegedűs was supported by the Bolyai János Scholarship of the Hungarian Academy of Sciences and the ÚNKP-19-4-SZTE-20 New National Excellence Program of the Ministry for Innovation and Technology.}}
\titlerunning{A Study of Security Vulnerability Types in Different Languages}
%
\author{G\'abor Antal\inst{1} \and 
Bal\'azs Mosolyg\'o\inst{1} \and
Norbert V\'andor\inst{1} \and
P\'eter Heged\H{u}s\inst{1,2}}
\authorrunning{Gábor Antal \and
Balázs Mosolygó \and 
Norbert Vándor \and
Péter Hegedűs
}
%
\institute{Department of Software Engineering, University of Szeged, Hungary
\and
MTA-SZTE Research Group on Artificial Intelligence, Szeged, Hungary }
\maketitle              
\begin{abstract}
The number of people accessing online services is increasing day by day, and with new users, comes a greater need for effective and responsive cyber-security.
Our goal in this study was to find out if there are common patterns within the most widely used programming languages in terms of security issues and fixes.
In this paper, we showcase some statistics based on the data we extracted for these languages.
Analyzing the more popular ones, we found that the same security issues might appear differently in different languages, and as such the provided solutions may vary just as much.

We also found that projects with similar sizes can produce extremely different results, and have different common weaknesses, even if they provide a solution to the same task.
These statistics may not be entirely indicative of the projects' standards when it comes to security, but they provide a good reference point of what one should expect.
Given a larger sample size they could be made even more precise, and as such a better understanding of the security relevant activities within the projects written in given languages could be achieved.


\end{abstract}

\section{Introduction}
\label{sec:intro}

As more and more vital services are provided by software systems accessible on the Internet, security concerns are becoming a top priority.
Mitigating the risks posed by malicious third parties should be at the core of the development processes.
However, eliminating all the security vulnerabilities is impossible, thus we have to be able to detect and understand the security issues in existing code bases.
How and what types of security vulnerabilities appear in programs written in various languages, and how their developers react to them are questions still lacking answers with satisfying empirical evidence.

In this paper, we present the results of a small-scale, open-source study that aims to show the differences between some languages based on their activity when it comes to fixing security issues.
We followed the basic ideas laid out by the work of Matt Bishop \cite{bishop2005introduction} with the design of our study approach.

We wanted to explore a set of patterns that could be later used as a point of reference. 
These are important not only when it comes to choosing the right language for a given task, but also to measure changes, improvements and deteriorations of the activity of the languages' communities.
To be able to derive meaningful conclusions, we investigated C, C++, BitBake, Go, Java, JavaScript, Python, Ruby and Scheme programs.
The choice to include so many languages had the advantage of not constraining our field of view to only certain kinds of projects.

For all the programs written in these different languages we extracted and analyzed the type of vulnerabilities found and fixed in the programs, the time it took for the fix to occur, the number of people working on a given project while an issue was active, and the required number of changes to the code and files to eliminate the issue.
In short, the results show that while the severity of an issue may correlate with the time it takes to fix it, that is not the case in general.
Averages show a similar pattern, which is likely because of the reintroduction of the same issues several times in larger projects.

CVEs (short for Common Vulnerabilities and Exposures) \cite{url_cve} are publicly disclosed cyber-security vulnerabilities and exposures that are stored and freely browsable online.
These can be categorized into CWEs (short for Common Weakness Enumeration) \cite{url_cwe}.
We used these entries to gauge the speed at which developers fix major issues in different programming languages.

We extracted our proxy metrics for vulnerabilities based on the textual analysis of git logs and as such may not be indicative of the actual development process.
Git is a free and open-source distributed version control system.\footnote{https://git-scm.com/}
Commits are a way to keep previously written code organized and available.
They usually have messages attached to them that explain what the contained changes are and what purpose do they serve.
We have used these messages to collect data about CVEs from commit messages.

We used a PostgreSQL database to store the collected CVE and CWE entries extracted from commit logs.
These were downloaded using an updated version of an open-source project called cve manager.\footnote{\url{https://github.com/aatlasis/cve\_manager}}
The commit messages were extracted using a mostly self-developed tool called git log parser.

We found that smaller and more user interface focused projects rarely document CVE fixes, however, larger-scale projects, especially those concerning backend solutions and operating systems (package managers, etc.) are more inclined to state major bug fixes.
We also found that in some projects, the developers prefer to only mention CVEs at larger milestones or releases, while in others, they were present in the exact commit they were fixed in.
The paper also looks at CWEs more specifically, their prevalence in different languages.
Some of these are language-specific, while others are more general.

\section{Approach}
\label{sec:approach}

Our main concern in this study was security, which led us to look for CVEs and CWEs in commit logs.
This is a good way to identify major and confirmed vulnerabilities without the need for in-depth code analysis.
We found that there are clear trends in some languages when it comes to handling various vulnerability types (CWEs).
These can help others to apply a solution for an issue since these statistics can serve as guides that show what to watch out for.

The approach we took can be best explained through the tools we created to collect the necessary information.
We will use the described tools as bullet points to illustrate the flow of the entire study and the inner workings of the miner.
In the approach summary, we will explain things in more detail and also explain the design decisions we took during planning the approach.

\subsection{CVE Manager}
CVE Manager\footnote{\url{https://github.com/gaborantal/cve\_manager}} is the backbone of most statistics and is essential to validating the found CVE entries. 
It is a lightweight solution that downloads the CVE data from the MITRE Corporation's\footnote{https://www.mitre.org/} website.
We store most of the collected data in a PostgreSQL\footnote{https://www.postgresql.org/about/} database.
The tool is used to query for CVE entries found by the miner and some of their properties like their \emph{id, impact score, severity}, and so on.

\subsection{Git Log Parser}
The other important tool used by the miner is our git log parser\footnote{\url{https://github.com/gaborantal/git-log-parser}} solution.
It simulates user commands using the Python subprocess module, which allows it to bypass some of the git's limitations.
The script is prepared to mine local directories for data in the contained repository's commits.
The parser first navigates to the path provided by the user through command line input, then issues the git log command that lists every commit and their meta-data. 

It then saves this information into a list that will later be printed into a JSON file.
This basic data is being extended with the line and file change information by comparing each commit to its predecessor with the git diff command.

The reports generated by the parser can be useful in a variety of situations, similar to ours, where an external utility needs the logs of a specific git repository.
Some of its results are not used by the miner but are intended for later use, for example, the parser could check whether a commit is a merge or not, which is currently ignored in finding CVEs.

\subsection{CVE Miner}
\begin{figure}
\begin{center}
\includegraphics[width=115mm]{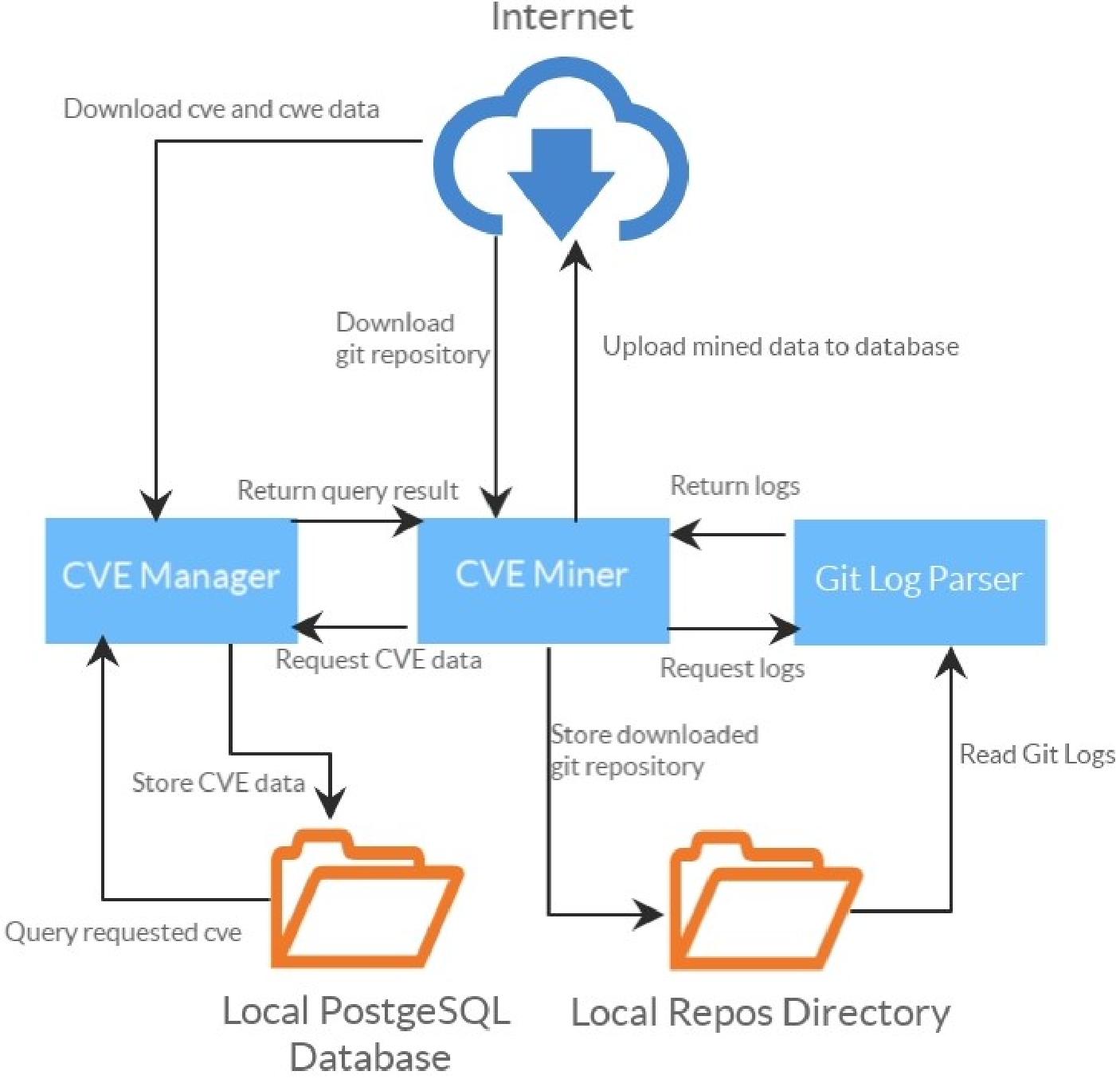}
\caption{A schematic representation of our miner}
\label{fig:fig1}
\end{center}
\end{figure}

The main tool of our project is the miner\footnote{\url{https://github.com/gaborantal/cve-miner}}, which uses both the CVE Manager and the Git Log Parser to create a JSON file and a database entry for each CVE found and presumably, fixed in the actual repository.
Figure~\ref{fig:fig1} represents the inner working of the miner and its interaction with the other tools.

The miner requires some initial setup since the CVE data needs to be downloaded and inserted into a local PostgreSQL database.
This is done in two steps.
In the first step, the data is collected into a local NVD directory from which we read and upload it to the database in the second step.

There are multiple ways to start working with the CVE Miner.
It can mine from both local and online sources.
These options can be accessed using the command-line interface.
When an online source is provided, a ``repos'' directory will be created if one does not already exist and the given repository will be automatically downloaded into it.
Then the miner will continue as if a local directory had been provided.
Multiple targets can be specified at once using a JSON file and the appropriate command-line argument.

The miner then processes the repositories by using Git Log Parser.
After the JSON file is generated, the tool searches the messages attached to the commits for CVE entries.
If a CVE is mentioned once, the miner assumes that the associated commit fixes the CVE.
If it is mentioned multiple times, it is assumed that the first occurrence implies that the CVE is found in the code, and any subsequent mentions are the fixes for that vulnerability.
During this process, other data is collected, including but not limited to the contributors, the number of changed files, and the number of commits between the finding and fixing of the CVE.

The next step is the calculation of statistics.
The miner uses the previously acquired information to calculate the average time between the commit that found the CVE and the commit that fixed it.
The other part of our statistics is correlation testing.
The tool calculates the correlation between a CVE entry's severity and the time needed to fix it.

The last step is storing the data.
By default, the miner creates a JSON file containing all the found CVEs and the calculated statistics.
If chosen, the tool also uploads it to an Airtable~\footnote{https://airtable.com/product} database.

\subsection{Approach Summary}
Our main point of interest during this study was the collection of security-related data, thus a large emphasis has been put on it.
We focused mainly on creating useful utilities for later research, and feel like we succeeded when it comes to most of the tools created.

We took the approach of looking only for mentions of the text ``CVE'' in commit logs as it is a fast solution providing sufficiently good approximation.
The best way to improve current data is of course to collect a much larger amount of them.

\section{Results}
\label{sec:results}

\subsection{Time Based Statistics}

\paragraph{Time elapsed between the finding and fixing commit.}

This statistic can be interpreted in multiple ways.
First, we will cover the intended purpose, showing how long it takes on average to fix a CVE entry.
This is more accurate on projects at a smaller scale or lifespan since those have a lower chance of false fixing claims and reoccurring issues.
Since we only check the textual references of CVE entries in commit logs not the actions taken, these properties of the projects are important.

\begin{figure}[H]
\centering
\captionsetup{justification=centering}
\includegraphics[width=85mm]{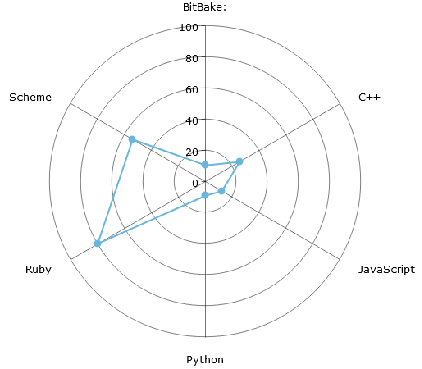}
\caption{The average time elapsed in days between finding and fixing a CVE}
\label{fig:fig2}
\end{figure}

The second way the statistics can be interpreted is, as we mentioned previously, an indication of reoccurring vulnerabilities.
Most of the time a CVE entry is mentioned in the context where it is claimed to be fixed, which is not surprising since you would not publicize an actual security issue in your system.
Based on this, most CVEs should be mentioned only once.
However, this is not the case with most large scale projects.
We hypothesize that this happens because later changes may reintroduce a vulnerability previously fixed, which is likely because in larger systems it is a lot harder to foresee every possible outcome a change might have.
Projects with longer code history usually have more reoccurring issues than others.
When it comes to languages, a similar pattern can be observed (see Figure~\ref{fig:fig2}).
The differences are drastic since the scale and age of the analyzed projects vary.
Most of the C++ and Scheme projects we looked at were larger projects, hence the reason for their dominance in the chart.
Ruby is an outlier, there it is common for an issue to resurface years after the vulnerability has been fixed.
The other reason vulnerabilities in some languages are more prevalent than in the others also has to do with the fact that larger systems usually do not allow developers to make changes directly to the working tree, merges that happen later can also increase this fixing time.
This is not a huge issue since an error being fixed in a branch should not be considered fixed in the application until it has not been merged.

\paragraph{Time elapsed between the publication and fixing of a CVE.}

\begin{figure}[H]
\centering
\captionsetup{justification=centering}
\includegraphics[width=85mm]{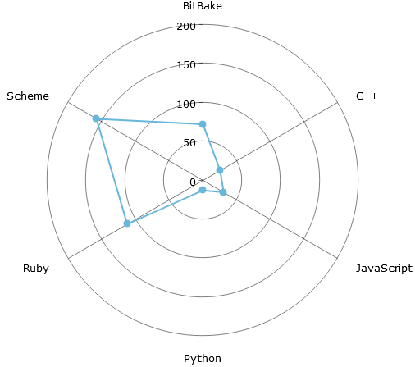}
\caption{The average time elapsed between the publication and fixing of a cve represanted in days}
\label{fig:from_pub}
\end{figure}

This statistic (see Figure~\ref{fig:from_pub}) is similar in nature to the previous one, however, it also takes into account the time each CVE spent in the code unnoticed after its publication.
Most of the languages show similar attributes compared to the previous chart, however, when it comes to BitBake, a clear bump is visible, implying that it takes longer to come up with the first fix for an issue in BitBake programs.

\paragraph{Correlation between time and severity.}

\begin{figure}[h]
\centering
\captionsetup{justification=centering}
\includegraphics[width=\textwidth]{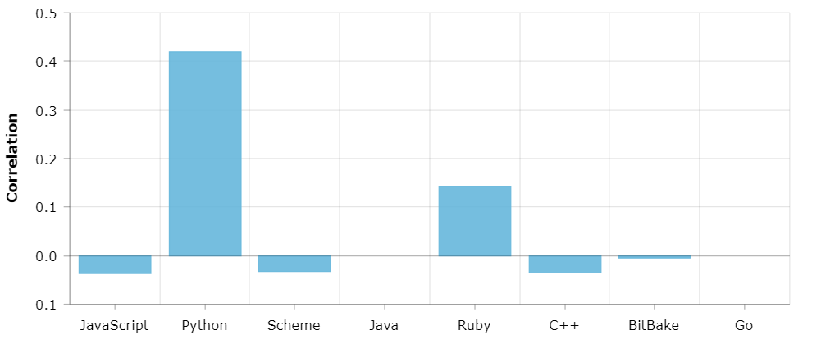}
\caption{The correlation between the base score(severity) and time taken fixing the cve}
\label{fig:correlations}
\end{figure}

The chart in Figure.~\ref{fig:correlations} shows the correlation between the aforementioned statistics and the severity of the CVEs.
The correlation between the fixing and the finding of a CVE can be attributed to the difficulty of the issue at hand, and the thoroughness of the testing.

The correlation between the publication date of CVEs and the time it took to fix them shows how prepared developers were when it came to fixing these vulnerabilities, since, for example, in the case of Python, the more severe problems were solved quicker than the others.
This might imply that they put a larger emphasis on getting rid of more severe issues.

\subsection{Activity Based Statistics}

\paragraph{Active contributors and commit count during the fixing of a CVE}

\begin{figure}[H]

\captionsetup{justification=centering}
\begin{center}
\includegraphics[width=1\textwidth]{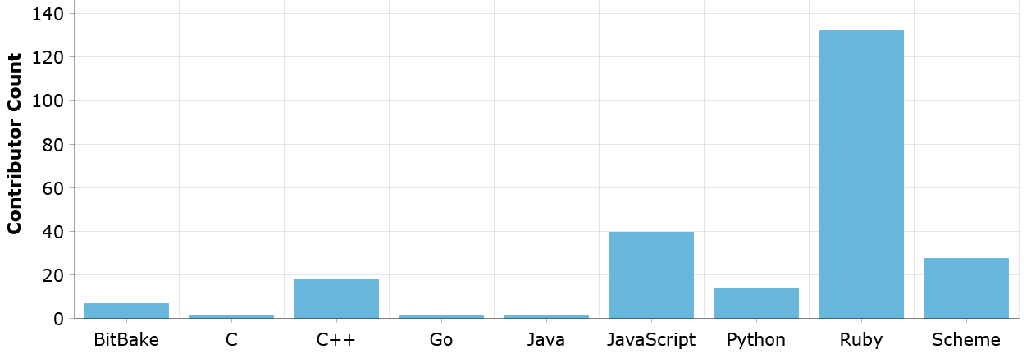}
\end{center}
\caption{The average number of contributors between the finding and fixing commit} \label{fig:contributor_count}

\end{figure}

\begin{figure}[H]
\centering
\captionsetup{justification=centering}
\includegraphics[width=\textwidth]{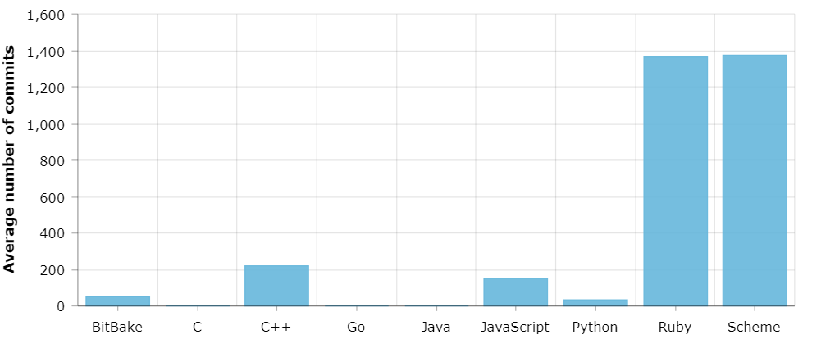}
\caption{The average number of commits between the finding and fixing of a CVE} \label{fig:commit_count}
\end{figure}

The results in Figures~\ref{fig:contributor_count} and \ref{fig:commit_count}) showcase not only how quickly some issues might be fixed, but the activity within the project during the process of fixing an issue.
Both charts show activities withing projects, Figure~\ref{fig:contributor_count} the number of contributors working on the code between the first and last commit mentioning the same CVE. As we can see several tens (e.g. JavaScript, Scheme) or even above 100 (Ruby) contributors might work on a codebase in the period of fixing a security vulnerability.

The number of commits in the vulnerability fixing period is highest in Ruby and Scheme (almost 1400).
This implicates that a lot of code changes happen while a security vulnerability is finally fixed.

\subsection{Average File and Line Changes}

The average changes to files and lines show how impactful an average CVE is in each language.
These numbers are of course extremely varied not only per language but per project as well since some might use fewer, but longer files, to store the same code, while others might separate code a bit more.
They also might only mention CVEs at larger milestones or merges, making some of the results disorderly high.
Table~\ref{table:avg_total_changes} shows the average total lines and files changed per language upon fixing a CVE.

\begin{table}
\begin{center}
\vspace{-20pt}
\caption{Average total changes}
\label{table:avg_total_changes}
\begin{tabular}{|c|c|c|}
\hline
Language & Average Total Lines Changed & Average Total Files Changed\\
\hline
\hline
BitBake &  {153.25} &  {2.5}\\
\hline
Go & {123.23} & {5.69}\\
\hline
JavaScript & {288.46} & {4.53}\\
\hline
Python & {84.16} & {5.7}\\
\hline
Ruby & {58.77} & {4.19}\\
\hline
\end{tabular}
\vspace{-20pt}
\end{center}
\end{table}

\subsection{Most Common CWEs by Language}

\paragraph{The usefulness of CWEs.}
CWEs are a grouping used for CVEs based on the weaknesses they cause.
Knowing which CWE is most common in a language can be extremely useful when it comes to finding, fixing, and looking out for problems.
This can reduce the time and energy needed to overcome certain vulnerabilities, and can raise the quality of code.

Our figures in Table~\ref{table:common_cwes} are of course not entirely indicative of each language as our scope is very limited, however, it might still give an idea of what to look for.
As an example, the most common CWE in C++ is CWE-119~\footnote{https://cwe.mitre.org/data/definitions/119.html} which has to do with incorrect memory management.
An example of a more general CWE is CWE-20~\footnote{https://cwe.mitre.org/data/definitions/20.html}, a possible cause of this is an improper input validation in the code.

\begin{table}[!htb]
  \centering
  \vspace{-30pt}
  \begin{minipage}[t]{.5\textwidth}
    \centering
    \begin{table}[H]
      \begin{center}
        \begin{tabular}{|c|c|c|}
          \hline
          Language   & CWE Group     & Percentage \\
          \hline
          \hline
          BitBake    & CWE-119       & 21.40\%    \\
                     & CWE-20        & 11.74\%    \\
                     & CWE-125       & 18.87\%    \\
          \hline
          C          & CWE-400       & 25.00\%    \\
                     & CWE-125       & 25.00\%    \\
                     & CWE-20        & 50.00\%    \\
          \hline
          C++        & CWE-119       & 92.39\%    \\
                     & CWE-200       & 5.71\%     \\
          \hline
          Java       & CWE-200       & 15.00\%    \\
                     & CWE-502       & 45.00\%    \\
                     & CWE-20        & 20.00\%    \\
          \hline
          JavaScript & CWE-119       & 6.45\%     \\
                     & NVD-CWE-Other & 5.38\%     \\
                     & CWE-20        & 13.98\%    \\
                     & CWE-400       & 15.05\%    \\
                     & CWE-200       & 12.90\%    \\
                     & CWE-79        & 7.53\%     \\
          \hline
        \end{tabular}
      \end{center}
    \end{table}
  \end{minipage}%
  \begin{minipage}[t]{0.5\textwidth}
    \centering
    \begin{table}[H]
      \begin{center}
        \begin{tabular}{|c|c|c|}
          \hline
          Language   & CWE Group     & Percentage \\
          \hline
          \hline
          Go         & CWE-400       & 25.00\%    \\
          \hline
          Python     & CWE-200       & 9.68\%     \\
                     & CWE-79        & 16.13\%    \\
                     & CWE-601       & 9.68\%     \\
                     & CWE-185       & 6.45\%     \\
                     & CWE-20        & 16.13\%    \\
                     & CWE-89        & 9.68\%     \\
          \hline
          Ruby       & CWE-79        & 26.92\%    \\
                     & CWE-20        & 15.38\%    \\
                     & CWE-264       & 11.54\%    \\
                     & CWE-89        & 9.62\%     \\
                     & CWE-22        & 5.77\%     \\
                     & CWE-200       & 5.77\%     \\
          \hline
          Scheme     & CWE-20        & 8.29\%     \\
                     & CWE-119       & 23.49\%    \\
                     & CWE-125       & 8.09\%     \\
                     & CWE-416       & 7.70\%     \\
          \hline
        \end{tabular}
      \end{center}
    \end{table}
  \end{minipage}
  \caption{Most Common CWEs per languages}
  \label{table:common_cwes}
  \vspace{-30pt}
\end{table}

For some of the languages with a more diverse set of CWEs, we created pie charts (see Figure~\ref{fig7}), to visually illustrate their distribution.
\vspace{-10pt}

\section{Related Work}
\label{sec:related}

There are plenty of previous works investigating different aspects of security vulnerabilities.

Li and Paxson \cite{li2017large} conducted a large-scale empirical study of security patches.
They investigated more than 4,000 bug fixes that affected more than 3,000 vulnerabilities in 682 open-source software projects.
They also used the National Vulnerability Database as a basis, but they used external sources (for example GitHub) to collect information about a security issue.
We only rely on data that provided by NVD \cite{url_nvd} or MITRE \cite{url_cve,url_cwe}.
In their work, they investigated the life-cycle of both security and non-security patches, compared their impact on the code base, their characteristics.
They found out that security patches have a lower footprint in code bases than non-security fixes; the third of all security issues were introduced more than 3 years before the fixing patch, and there were also cases when a security bugfix failed to fix the corresponding security issue.

Frei et al. \cite{frei2006large} presented a large-scale analysis of vulnerabilities, mostly concentrated on discovery, disclosure, exploit, and patch dates.
The authors have found out that until 2006, the hackers reacted faster to vulnerability than the vendors. 

Similar to the previous work, Shahzad et al. \cite{shahzad2012alarge} presented a large-scale study about various aspects of software vulnerabilities during their life cycle.
They created a large software vulnerability data set with more than 46,000 vulnerabilities.
The authors also identified the most exploited forms of vulnerabilities (for example DoS, XSS).
In our research, we also use categories, however, our categories are defined by CWE.
They found that since 2008, the vendors have become more agile in patching security issues.
They also validated the fact the vendors are getting faster than the hackers since than.
Moreover, patching of vulnerabilities in closed-source software is faster than open-source software.

Kuhn et al. \cite{kuhn2017analysis} analyzed the vulnerability trends between 2008 and 2016.
They also analyzed the severity of the vulnerabilities as well as the categories.
They found that number of design-related vulnerabilities is growing while there are several other groups (for example CWE-89 (SQL Injection)) that show a decreasing trend.

In their work, Wang et al. used Bayesian networks to categorize CVEs.
They used the vulnerable product and CVSS~\footnote{Common Vulnerability Scoring System, as presented by Mell et al.~\cite{mell2006common}} base metric scores as the observed variables.
Although we do not use any machine learning methods in this study, our long term goal is to use various machine learning methods using the data presented in this study.
Wang et al. proved that categorizing CVEs is possible and machine learning can do that.

Gkortzis et al. presented VulinOSS, a vulnerability data set containing the vulnerable open-source project versions, the details about the vulnerabilities, and numerous metrics related to their development process (e.g. whether they have tests, static code metrics).

In their work, Massacci et al. analyzed several research problems in the field of vulnerability and security analysis, the corresponding empirical methods, and vulnerable prediction.
They summarized the databases used by several studies and identified the most common features used by researchers.
They also conducted an experiment in which they integrated several data sources on Mozilla Firefox.
The authors also showed that different data sources might lead to different results to a specific question.
Therefore, the quality of the database is a key component.
In our paper, we try our best to provide good quality and usable database for further researches.

Abunadi et al.~\cite{abunadi2015towards} presented an empirical study aiming to clarify how useful cross-project vulnerability prediction could be.
They conducted their research on a publicly available data set in the context of cross-project vulnerability prediction.
In our research, we collected data from several programming languages.
Hence we believe that our data set can be used in cross-project vulnerability prediction. 

Xu et al.~\cite{xu2017spain} presented a low-level (binary-level) patch analysis framework, that can identify security and non-security related patches by analyzing the binaries.
Their framework can also detect patterns that help to find similar patches/vulnerabilities in the binaries.
In contrast to their work, we use data mining and static process metrics.
Therefore, our approach does not need any binaries, it does not require the project to be in an executable state which can be extremely useful when a project's older version could not be compiled.

V{\'a}squez et al.~\cite{Vsquez2017AnES} analyzed more than 600 Android-related vulnerabilities and the corresponding patches.
Their approach uses NVD and Google Android security bulletins to identify security issues. 
Despite that we do not include Android security bulletins in this research, we plan to extend our scope in the future and include those vulnerabilities too as our framework is extensible.

Identifying whether a change contains security fix or not can be also challenging ~\cite{sliverski2005when,10.1145/1083142.1083147}.

In our paper, we use the data of a vulnerability and we then we find the corresponding commits for a vulnerability.

Vaidya et al. \cite{vaidya2019security} analyzed two language-based software ecosystems in the aspect of security.
They investigated npm's and PyPI's ecosystem and some of the recent security attacks.
They found out that automated detection of malicious packages is unfeasible, but using tools and metrics might help.
In our work, we are providing some of the metrics and data that can help in detecting malicious commits.

In order to improve the quality of a software system, one has to evaluate the software's quality. 
This can be done in several ways, for example we can use data mining, textual analysis or we can estimate the software's quality and/or reliability.
Some works uses machine learning  \cite{behera2018software,shukla2018software} in order to capture the different characteristics of a software.
That can also help to find vulnerable components.

Rahimi and Zargham proposed a method \cite{rahimi2013vulnerability} to automatically predict vulnerability discovery in softwares.
We believe that our data can be useful in learning models like the previously mentioned vulnerability discovery model.

Several works use bug reports to identify bugs and security issues in code bases \cite{wu2011bugminer}, \cite{wijayasekara2012mining}.

In their work, Neuhaus et al. \cite{10.1145/1315245.1315311} use existing vulnerability database to mine vulnerability data and use the collected data to predict whether a given software component is likely to contain a vulnerability or not.
Li et al. proposed \cite{8026953} a vulnerability mining algorithm that also uses CVE, CWE data sets in order to mine the vulnerabilities.
In contract to their work, we rely on only already fixed vulnerabilities that has a remark on the source code (and the corresponding version control system).

In their work, Gyimesi et al. \cite{GVS19} uses GitHub's issue management tools to find bugs and the corresponding code snippets.
In contrast to their work, In contract to their work, we rely on only already fixed vulnerabilities that has a remark on the source code (and the corresponding version control system).

In our work, we did not try to reuse any of the existing bug databases as  Munaiah et al. proved that there is only a weak correlation between number of bugs and number of vulnerabilites \cite{Munaiah2016DoBF,7180086} in softwares.

\section{Threats to Validity}
\label{sec:threats}

The main weakness of our results is the limited scale at which we operated.
We only had the resources to mine a few repositories for most languages.
For this reason, some tables and graphs are missing some languages as one major project's practices had too large of an effect on the overall statistics.

One other major issue stems from the fact that we do not look at the code, but rely on the commit messages left by the developers.
This can be troublesome when it comes to claiming that issues reappear since it could be the case that they were never fixed in the first place.
The way we check for CVE fixes is also fairly limited, since we only look for the mention of a CVE in the commit message, but do not check the context in which it appears.
We assume that the last commit at which a CVE is mentioned is the last time it occurred, and has therefore been fixed.
This might not be the case, it is possible that later a fix happened, but the developer forgot to mention it.

We also do not account for merges, which can increase the number of lines needed for a fix.
We believe that an issue is not fixed until it is merged into the master branch.
However, if we count the lines in the commit that fixed the issue and the number of lines present in a merge that contains the commit getting rid of a vulnerability might not be indicative of the actual amount of work needed for a solution.
In these cases, we currently just count the lines twice, but this has caused some statistics to be left out since they portrayed false information because of the practices the developers used when they merged larger pieces of code at once.

\section{Conclusion}
\label{sec:conclusion}

We presented a study that focuses on security issues.
Our main goals were to determine if there were vulnerability types characteristic of languages.
More specifically, what these issues are, how quickly they get fixed, and how efficiently does that fix happen.

We found that even at smaller sample sizes, specific weaknesses showed a clear trend in most of the tested languages.
For example, in the case of C++ CWE-119 (memory handling problems) was the biggest group of issues faced by developers.
This may not surprise those familiar with the language, but for a new developer, it can be a clear pointer as to what to watch out for.

The best example of how interesting these statistics truly are is Ruby.
It is visible that for Ruby developers, the biggest issue is CWE-79, improper neutralization of inputs.~\footnote{https://cwe.mitre.org/data/definitions/79.html}
These issues take less effort to fix than others, requiring on average about 60 lines of code changes, and 4 file changes, however, the same issue might reappear later, as shown in Figure~\ref{fig:fig2}.
It is also visible that while in Ruby it takes the least amount of lines to fix an issue, more severe vulnerabilities take longer to get rid of, as seen in Figure~\ref{fig:correlations}.

In conclusion, each language has its share of common weaknesses, which depend on a variety of factors, and being cautious of these is important.

\bibliographystyle{splncs04}
\bibliography{bibl}

\begin{thebibliography}{10}
\providecommand{\url}[1]{\texttt{#1}}
\providecommand{\urlprefix}{URL }
\providecommand{\doi}[1]{https://doi.org/#1}

\bibitem{abunadi2015towards}
Abunadi, I., Alenezi, M.: Towards cross project vulnerability prediction in
  open source web applications. In: Proceedings of the The International
  Conference on Engineering MIS 2015. ICEMIS ’15, Association for Computing
  Machinery, New York, NY, USA (2015). \doi{10.1145/2832987.2833051},
  \url{https://doi.org/10.1145/2832987.2833051}

\bibitem{behera2018software}
Behera, R.K., Shukla, S., Rath, S.K., Misra, S.: Software reliability
  assessment using machine learning technique. In: International Conference on
  Computational Science and Its Applications. pp. 403--411. Springer (2018)

\bibitem{bishop2005introduction}
Bishop, M.: Introduction to computer security, vol.~50. Addison-Wesley Boston
  (2005)

\bibitem{7180086}
{Camilo}, F., {Meneely}, A., {Nagappan}, M.: Do bugs foreshadow
  vulnerabilities? a study of the chromium project. In: 2015 IEEE/ACM 12th
  Working Conference on Mining Software Repositories. pp. 269--279 (2015)

\bibitem{frei2006large}
Frei, S., May, M., Fiedler, U., Plattner, B.: Large-scale vulnerability
  analysis. In: Proceedings of the 2006 SIGCOMM workshop on Large-scale attack
  defense. pp. 131--138 (2006)

\bibitem{kuhn2017analysis}
Kuhn, D., Raunak, M., Kacker, R.: An analysis of vulnerability trends,
  2008-2016. pp. 587--588 (07 2017). \doi{10.1109/QRS-C.2017.106}

\bibitem{li2017large}
Li, F., Paxson, V.: A large-scale empirical study of security patches. In:
  Proceedings of the 2017 ACM SIGSAC Conference on Computer and Communications
  Security. pp. 2201--2215 (2017)

\bibitem{8026953}
{Li}, X., {Chen}, J., {Lin}, Z., {Zhang}, L., {Wang}, Z., {Zhou}, M., {Xie},
  W.: A mining approach to obtain the software vulnerability characteristics.
  In: 2017 Fifth International Conference on Advanced Cloud and Big Data (CBD).
  pp. 296--301 (2017)

\bibitem{mell2006common}
{Mell}, P., {Scarfone}, K., {Romanosky}, S.: Common vulnerability scoring
  system. IEEE Security Privacy  \textbf{4}(6),  85--89 (2006)

\bibitem{url_cve}
{MITRE Corporation}: {CVE - Common Vulnerabilities and Exposures}.
  \url{https://cve.mitre.org/} (2020), [Online; accessed 29-April-2020]

\bibitem{url_cwe}
{MITRE Corporation}: {CWE - Common Weakness Enumeration}.
  \url{https://cwe.mitre.org/} (2020), [Online; accessed 29-April-2020]

\bibitem{Munaiah2016DoBF}
Munaiah, N., Camilo, F., Wigham, W., Meneely, A., Nagappan, M.: Do bugs
  foreshadow vulnerabilities? an in-depth study of the chromium project.
  Empirical Software Engineering  \textbf{22},  1305--1347 (2016)

\bibitem{10.1145/1315245.1315311}
Neuhaus, S., Zimmermann, T., Holler, C., Zeller, A.: Predicting vulnerable
  software components. In: Proceedings of the 14th ACM Conference on Computer
  and Communications Security. p. 529–540. CCS ’07, Association for
  Computing Machinery, New York, NY, USA (2007). \doi{10.1145/1315245.1315311},
  \url{https://doi.org/10.1145/1315245.1315311}

\bibitem{GVS19}
P{\'e}ter, G., B{\'e}la, V., Andrea, S., Davood, M., {\'A}rp{\'a}d, B., Rudolf,
  F., Ali, M.: {BugsJS: A Benchmark of JavaScript Bugs}. In: Proceedings of the
  12th IEEE Conference on Software Testing, Validation and Verification (ICST).
  pp. 90--101. IEEE (Apr 2019). \doi{10.1109/ICST.2019.00019}

\bibitem{rahimi2013vulnerability}
{Rahimi}, S., {Zargham}, M.: Vulnerability scrying method for software
  vulnerability discovery prediction without a vulnerability database. IEEE
  Transactions on Reliability  \textbf{62}(2),  395--407 (2013)

\bibitem{shahzad2012alarge}
{Shahzad}, M., {Shafiq}, M.Z., {Liu}, A.X.: A large scale exploratory analysis
  of software vulnerability life cycles. In: 2012 34th International Conference
  on Software Engineering (ICSE). pp. 771--781 (2012)

\bibitem{shukla2018software}
Shukla, S., Behera, R.K., Misra, S., Rath, S.K.: Software reliability
  assessment using deep learning technique. In: Towards Extensible and
  Adaptable Methods in Computing, pp. 57--68. Springer (2018)

\bibitem{sliverski2005when}
Sliwerski, J., Zimmermann, T., Zeller, A.: When do changes induce fixes? In:
  Proceedings of the 2005 International Workshop on Mining Software
  Repositories. p. 1–5. MSR ’05, Association for Computing Machinery, New
  York, NY, USA (2005). \doi{10.1145/1083142.1083147},
  \url{https://doi.org/10.1145/1083142.1083147}

\bibitem{10.1145/1083142.1083147}
\'Sliwerski, J., Zimmermann, T., Zeller, A.: When do changes induce fixes? In:
  Proceedings of the 2005 International Workshop on Mining Software
  Repositories. p. 1–5. MSR ’05, Association for Computing Machinery, New
  York, NY, USA (2005). \doi{10.1145/1083142.1083147},
  \url{https://doi.org/10.1145/1083142.1083147}

\bibitem{url_nvd}
{U.S. National Institute of Standards and Technology}: {National Vulnerability
  Database}. \url{https://nvd.nist.gov/home} (2020), [Online; accessed
  29-April-2020]

\bibitem{vaidya2019security}
Vaidya, R.K., De~Carli, L., Davidson, D., Rastogi, V.: Security issues in
  language-based sofware ecosystems. arXiv preprint arXiv:1903.02613  (2019)

\bibitem{Vsquez2017AnES}
V{\'a}squez, M.L., Bavota, G., Escobar-Velasquez, C.: An empirical study on
  android-related vulnerabilities. Proceedings of the IEEE/ACM 14th
  International Conference on Mining Software Repositories (MSR) pp. 2--13
  (2017)

\bibitem{wijayasekara2012mining}
{Wijayasekara}, D., {Manic}, M., {Wright}, J.L., {McQueen}, M.: Mining bug
  databases for unidentified software vulnerabilities. In: 2012 5th
  International Conference on Human System Interactions. pp. 89--96 (2012)

\bibitem{wu2011bugminer}
Wu, L.L., Xie, B., Kaiser, G.E., Passonneau, R.: Bugminer: Software reliability
  analysis via data mining of bug reports  (2011)

\bibitem{xu2017spain}
{Xu}, Z., {Chen}, B., {Chandramohan}, M., {Liu}, Y., {Song}, F.: Spain:
  Security patch analysis for binaries towards understanding the pain and
  pills. In: Proceedings of the IEEE/ACM 39th International Conference on
  Software Engineering (ICSE). pp. 462--472 (May 2017).
  \doi{10.1109/ICSE.2017.49}

\end{thebibliography}

\end{document}